\documentclass{article}

\usepackage{microtype}
\usepackage{graphicx}
\usepackage{subfigure}
\usepackage{booktabs} 
\usepackage{graphics}
\usepackage{caption}
\usepackage[table]{xcolor}

\usepackage{hyperref}

\usepackage[accepted]{synsml2023}

\usepackage{amsmath}
\usepackage{amssymb}
\usepackage{mathtools}
\usepackage{amsthm}

\usepackage[capitalize,noabbrev]{cleveref}

\theoremstyle{plain}

\theoremstyle{definition}

\theoremstyle{remark}

\usepackage[textsize=tiny]{todonotes}

\synsmltitlerunning{Neural Modulation Fields for Conditional Cone Beam Neural Tomography}
\newcommand{\ratiominifigure}{.152}
\begin{document}

\twocolumn[
\synsmltitle{Neural Modulation Fields for Conditional Cone Beam Neural Tomography}



\synsmlsetsymbol{equal}{*}

\begin{synsmlauthorlist}
\synsmlauthor{Samuele Papa}{uva,nki}
\synsmlauthor{David M. Knigge}{uva,nki}
\synsmlauthor{Riccardo Valperga}{uva}
\synsmlauthor{Nikita Moriakov}{nki}
\synsmlauthor{Miltos Kofinas}{uva}
\synsmlauthor{Jan-Jakob Sonke}{nki}
\synsmlauthor{Efstratios Gavves}{uva}
\end{synsmlauthorlist}

\synsmlaffiliation{uva}{Institute for Informatics, University of Amsterdam, Amsterdam, the Netherlands}
\synsmlaffiliation{nki}{Netherlands Cancer Institute, Amsterdam, the Netherlands}

\synsmlcorrespondingauthor{Samuele Papa}{s.papa@uva.nl}
\synsmlcorrespondingauthor{David M. Knigge}{d.m.knigge@uva.nl}

\synsmlkeywords{Machine Learning}

\vskip 0.3in
]



\printAffiliationsAndNotice{\synsmlEqualContribution} 

\begin{abstract}
Conventional Computed Tomography (CT) methods require large numbers of noise-free projections for accurate density reconstructions, limiting their applicability to the more complex class of Cone Beam Geometry CT (CBCT) reconstruction. 
Recently, deep learning methods have been proposed to overcome these limitations, with methods based on neural fields (NF) showing strong performance, by approximating the reconstructed density through a continuous-in-space coordinate based neural network. Our focus is on improving such methods, however, unlike previous work,
which requires training an NF from scratch for each new set of projections, we instead propose to leverage anatomical consistencies over different scans by training a single \textit{conditional} NF on a dataset of projections. We propose a novel conditioning method where \textit{local} modulations are modeled per patient as a field over the input domain through a Neural Modulation Field (NMF). The resulting Conditional Cone Beam Neural Tomography (CondCBNT) shows improved performance for both high and low numbers of available projections on noise-free and noisy data.
\end{abstract}%
\begin{figure*}
  \centering \includegraphics[width=.822\textwidth]{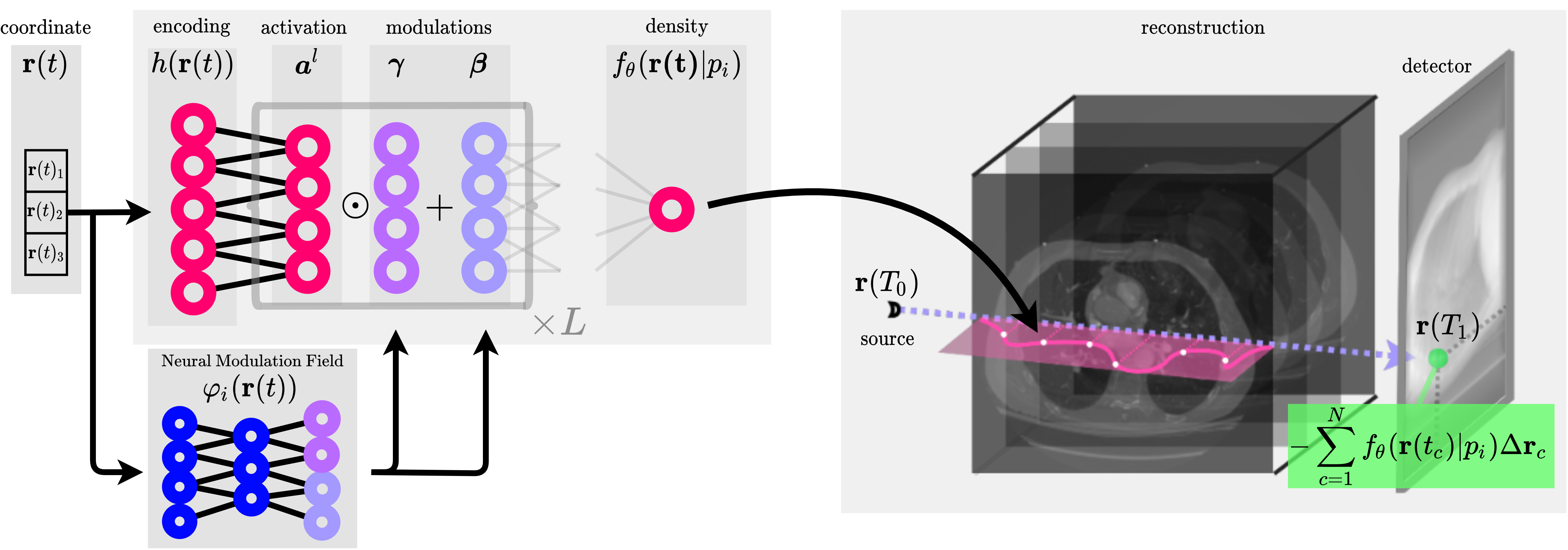}
  \caption{We propose \textit{Conditional Cone Beam Neural Tomography} (CondCBNT), a framework for reconstructing Cone Beam Computed Tomography volumes using neural fields. An integral is taken over values sampled from a neural field $f_{\theta}$ at coordinates $\mathbf{r}(t)$ along a ray cast from source to sensor. The coordinates are encoded into a multiresolution hash-encoding $h(\mathbf{r}(t))$ \cite{muller2022instant}, and passed through $L$ linear layers. To leverage consistencies over anatomies of different patients, we propose to model the density for a specific patient $p_i$ using a shared neural field $f_{\theta}$, whose activations $\boldsymbol{a}^l$ are modulated by a patient-specific \textit{Neural Modulation Field} (NMF) $\varphi_i$. This conditioning function learns a field of $\boldsymbol{\gamma}, \boldsymbol{\beta}$ FiLM modulations \cite{dumoulin2018feature} over the input space $\mathbb{R}^3$ for a patient $p_i$. The integral $-\sum_{c=1}^{N}f_{\theta}(\mathbf{r}(t_c))\Delta \mathbf{r}_c$ is supervised at the sensor using the corresponding observed projection value.\looseness=-1 }
  \label{fig:condcbnt}
\end{figure*}
\section{Introduction}
In inverse problems, the goal is to infer a certain quantity of interest from indirect observations. They arise in many scientific fields, medical imaging \cite{louisMedicalImagingState1992}, biology \cite{karwowskiInverseProblemsQuantum2009, sridharanModernMachineLearning2022}, and physics \cite{romanovInverseProblemsMathematical2018, theeventhorizontelescopecollaborationFirstM87Event2019}. Unfortunately, many inverse problems are inherently \emph{ill-posed}, i.e., there exist multiple solutions that agree with the measurements and these do not necessarily depend continuously on the data \citep{kabanikhin2008definitions}. These issues warrant further study
, and tools from machine learning and deep learning in particular have attracted a lot of attention recently.

In this work, we focus on Computed Tomography (CT) \cite{oldendorfQuestImageBrain1978}, a medical imaging technique for reconstructing material density\footnote{To be precise, we try to find the \textit{attenuation coefficients}, but we may use density interchangeably, as they are strongly related under assumptions that hold in our setting.} inside a patient, using the mathematical and physical properties of X-ray scanners. In CT, several X-ray scans--or \textit{projections}--of the patient are acquired from various angles using a \emph{detector}. 
An important variant of CT is Cone Beam CT (CBCT), which uses flat panel detectors to scan a large fraction of the volume in a single rotation. 
Unfortunately, CBCT reconstruction is harder in comparison to classical (helical) CT. This is caused by the inherent mathematical difficulty of Radon Transform inversion in the three-dimensional setting \cite{tuy1983}, physical limits of the detector, and characteristics of the measurement process such as noise. Traditional reconstruction methods include FDK \cite{feldkamp1984practical}, and iterative reconstruction \cite{Kaipio2005}. FDK filters the projections and applies other simple corrections to properly account for the physical geometry of the acquisition system. Iterative methods use optimization to find the density that most closely resemble the measurements once projected using a forward operator. 
In addition, deep learning has seen increasing use in the field, with algorithms such as learned primal-dual \cite{adlerLearnedPrimaldualReconstruction2018}, invertible learned primal-dual \cite{rudzusika2021invertible} and LIRE \cite{moriakov2022lire}.\looseness=-1

Recently, reconstruction methods that employ Neural Fields (NFs) have been proposed. \textit{NFs are a class of neural architectures that parameterize a field $f: \mathbb{R}^d \rightarrow \mathbb{R}^n$, i.e. a quantity defined over spatial and/or temporal coordinates, using a neural network $f_\theta$} (see \citet{xie2022neural} for a survey on NFs). In CT reconstruction, these architectures have been used to approximate the density directly over the volume space $\mathbb{R}^3$ \cite{intratomo, zha2022naf, lin2023learning}.
\citet{zha2022naf} proposed Neural Attenuation Fields (NAF), an approach to supervise NFs using only the measured attenuated photon counts at the detector.
Despite showing promising results, this method requires training a NF from scratch for each volume, prohibiting transfer of learned features across volumes through weight sharing. Instead, \citet{lin2023learning} propose encoding a set of projections into a latent space shared over all training volumes, and decoding this into a density modeled as a NF.
However, encoding of all available projections is only feasible when a small number of them is used, as it would otherwise result in prohibitive compute and memory requirements. \looseness=-1 

In this work, we instead aim to remove the need for an explicit decoder. We leverage the work of \citet{park2019deepsdf}, who propose to learn latent codes for a dataset of 3D shapes using \textit{auto-decoding}, where randomly initialized latent codes are optimized during training. \citet{dupont2022data} expand on by using these learned latent codes as modulations for a shared NF. 
\citet{bauer2023spatial} show that the use of a single global code per signal limits reconstruction quality, and instead use a spatially structured grid of codes. Their approach greatly increases reconstruction quality, but requires interpolating a grid of modulations, increasing computational requirements for signals over higher-dimensional domains. 
We introduce the \textbf{Neural Modulation Field} (NMF) which models a continuous field of modulations over the signal domain. We propose the \textbf{Conditional Cone Beam Neural Tomography} (CondCBNT) framework, which incorporates this \textit{local conditioning function} to speed up reconstruction, while still processing all available projections, relieving restrictions on projection counts used in the reconstruction process. In doing so, we show considerable improvements in scenarios with both sufficient or limited projections, as well as in the presence of both noisy and noise-free data. \looseness=-1 
\section{Method}
Beer-Lambert's law relates the attenuation of electromagnetic radiation such as visible light or X-rays to the properties of the material it is traveling through \cite{swinehartBeerLambertLaw1962}. Let $\mathbf{r}:[T_0, T_1] \longrightarrow \mathbb{R}^{3}$ be the straight path taken by radiation through the medium. The radiation intensity $I(\mathbf{r}(T_1))$ at position $\mathbf{r}(T_1)$ is the line integral
\begin{equation}\label{eq:attenuation}
I(\mathbf{r}(T_1)) = I_0\exp{\left[-\int_{T_0}^{T_1}\mu(\mathbf{r}(t)) \, \bigl\lvert\mathbf{r}'(t)\bigr\rvert \, dt\right]},
\end{equation}
where $\mu:\mathbb{R}^3 \longrightarrow \mathbb{R}^{+}$ is the attenuation coefficient of the medium and $I_0 = I(\mathbf{r}(T_0))$ is the initial intensity.
The integral in \eqref{eq:attenuation} can be approximated by the sum
\begin{equation}\label{eq:sum}
    I(\mathbf{r}(T_1)) \approx I_0\exp{\left[-\sum_{c=1}^{N}\mu(\mathbf{r}(t_c))\, \bigl\lvert\mathbf{r}'(t_{c})\bigr\rvert \, \Delta t\right]},
\end{equation}
where $t_c\in [T_0, T_1]$ and $\left|\mathbf{r}'(t_{c})\right|\Delta t = \Delta \mathbf{r}_c = \left|\mathbf{r}(t_{c+1}) - \mathbf{r}(t_{c})\right|$.
Given a set of 2D CBCT projections $v_{\alpha} \in \mathbb{R}^{H \times W}$ with $H, W$ the height and width of the sensor and $\alpha$ the angle under which the projection was taken, we are trying to estimate density values along rays cast from source to sensor.
Each ray is the straight path $\mathbf{r}$ which connects the source to pixels in the detector. For simplicity, we bound the patient volume with a box and assume zero attenuation outside the box. Therefore, for every path, we compute the sum in \eqref{eq:sum} with only those $\mathbf{r}(t_c)$ that are contained in the bounding box. By taking the logarithm we can avoid the computationally tedious exponential and use $\log{I(\mathbf{r}(T_1))} \approx -\sum_{c=1}^{N}\mu(\mathbf{r}(t_c))\Delta \mathbf{r}_c + \log{I_0}$ and discard the constant that depends on the initial intensity, which we assume is the same for all projections. We use a neural field $f_\theta : \mathbb{R}^{3} \longrightarrow \mathbb{R}^{+}$ to approximate the density $\mu$ such that the intensity $I(\mathbf{r}(T_1))$ coincides with the intensity recorded by the detector at the position $\mathbf{r}(T_1)$: 
\begin{equation}
    \log{I(\mathbf{r}(T_1))} \approx -\sum_{c=1}^{N}f_{\theta}(\mathbf{r}(t_c))\Delta \mathbf{r}_c.
\end{equation}
\paragraph{Coordinate embedding.} \citet{tancik2020fourier} showed that ReLU MLPs suffer from spectral bias, limiting their capacity to model high frequency functions on low-dimensional domains. As a solution, they note that it is possible to embed coordinates $\mathbf{r}(t_c) \in \mathbb{R}^3$ into a higher-dimensional space $\mathbb{R}^{e}$ with $e\gg 3$ before passing them through the MLP. We choose to follow \citet{muller2022instant} and use the \textit{multiresolution hash-encoding}, denoted $h(\mathbf{r}(t_i))$, as it empirically shows fastest convergence in our experiments. See Appx. \ref{appx:hashencoding} for a full description of this embedding.
\paragraph{Conditioning with Neural Modulation Fields.} \label{par:conditioning}
Conditioning in neural fields consists of modulating the weights $\theta$ or activations $\boldsymbol{a}$ of a NF $f_{\theta}$ with a conditioning variable $\boldsymbol{z}$ to vary the NF's output \cite{xie2022neural}, a method often used to encode different samples $x_i$ from a single dataset $X$ through a set of latents $\{\boldsymbol{z}_i|x_i \in X\}$. Intuitively, in the setting of CT reconstruction, we could fairly assume the densities for patients $p_i \in P$ share a lot of anatomical structure. A conditional NF that is tasked with reconstructing a dataset of multiple volumes would be able to leverage this consistency in anatomical information in its reconstruction (e.g. inferring from noisy or missing data), with patient-specific characteristics being refined with the conditioning variable $\boldsymbol{z}_i$. To this end, we could in principle use the aforementioned auto-decoding approach with a \textit{global} conditioning latent $\boldsymbol{z}_i$. However, global conditioning has been shown to result in reconstructions with limited detail \cite{dupont2022data, bauer2023spatial}. This limitation is significant because patient-specific fine-grained details in scans contain information crucial for medical purposes.
We instead opt for \textit{local} conditioning, where the conditioning variable $\boldsymbol{z}_i$ depends on the input coordinate $\boldsymbol{r}(t)$. In previous works, this is done through interpolation of a trainable discrete data structure, e.g. a grid of latent codes \cite{shaham2021spatially, yu2021pixelnerf, bauer2023spatial}. Instead, to further increase expressivity of the resulting modulation and forego modelling choices such as code grid resolution and interpolation method, we propose to abstract the learning of modulations away from a discrete data structure and model the modulations themselves as a continuous field through a patient-specific \textit{Neural Modulation Field} (NMF) we denote $\varphi_i$. 
During training, parameters $\theta_i$ of the patient-specific NMFs $\varphi_{\theta_i}$ are optimized alongside the weights of the shared NF $f_{\theta}$, during inference - for a novel set of projections - only the parameters for $\theta_i$ are optimized.

For the activation modulation, we use feature-wise linear modulations (FiLM) \cite{dumoulin2018feature}, such that activations $\boldsymbol{a}^l$ at a layer $l$ with weights $\boldsymbol{W}^l$ and bias $\boldsymbol{b}^l$ are transformed with patient-specific $\textit{local}$ scaling and shifting modulations $\boldsymbol{\gamma}_i, \boldsymbol{\beta}_i$, as follows:
\begin{equation}
    \boldsymbol{a}^l_i = \operatorname{ReLU}((\boldsymbol{W}^l\boldsymbol{a}^{l-1}_i + \boldsymbol{b}^{l}) \odot \boldsymbol{\gamma}_i + \boldsymbol{\beta}_i),
\end{equation}
where $\boldsymbol{\gamma}_i, \boldsymbol{\beta}_i$ are obtained from the NMF $\varphi_{\theta_i}:\mathbb{R}^3\rightarrow \mathbb{R}^{\operatorname{dim}(\boldsymbol{\gamma})+\operatorname{dim}(\boldsymbol{\beta})}$. For specific architectural choices of the NMF and shared NF, see Appx. \ref{appx:experimental_details}. We term the resulting model \textit{Conditional Cone Beam Neural Tomography} (CondCBNT). See Fig. \ref{fig:condcbnt} for an overview of the framework.
\paragraph{Dataset.} The dataset used is derived from the LIDC-IDRI \cite{armato2015data}. This is a collection of diagnostic lung cancer screening thoracic CT scans. A random selection of $250$ cases was chosen and the CT scan resampled to $2$mm resolution. Then, each volume is projected using $256\times 256$ pixel, $2$mm resolution detectors. Angles equally spaced between $0^{\circ}$ and $205^{\circ}$ are used. $400$ projections are created, first without any noise, then with Poisson noise, used to simulate measurement noise with $5\times 10^5$ photons. A subset of $50$ equally-spaced projections is obtained from both. The $250$ volumes are split into $200/25/25$ for training, validation, and testing. The resulting dataset will be made publicly available upon acceptance. \looseness=-1
\paragraph{Metrics.}
For quantitve evaluation we rely on the \textit{Peak Signal to Noise Ratio} (\textbf{PSNR}), a classical measure of signal quality, and the \textit{Structural Similarity Index Measure} (\textbf{SSIM}), which captures the perceptive similarity between two images by analyzing small local chunks \cite{ssim}. Historically, both metrics have been defined for images, but we compute them over full volumes. Finally, we track the GPU memory used and the time required to reconstruct a volume.\looseness=-1
\paragraph{Baselines.}
FDK reconstruction \cite{feldkamp1984practical} was performed using Operator Discretization Library \cite{odl2017}. As an iterative reconstruction baseline, we implemented Landweber iteration with Total Variation regularization \cite{Kaipio2005}, where parameters such as step size, iteration count and the amount of regularization were chosen via grid search on the validation set. As a deep learning reconstruction baseline, we use LIRE-32(L) architecture from \citet{moriakov2022lire}, which is a dedicated lightweight, memory-efficient variant of learned primal-dual method from \citet{adlerLearnedPrimaldualReconstruction2018} for CBCT reconstruction. From the NF class of models, we compare with \citet{zha2022naf};
we do not compare with \citet{lin2023learning} due to their prohibitive computational costs.
\begin{table}[t]
    \begin{minipage}{0.48\textwidth}
    \caption{Mean $\pm$ standard deviation of metrics over test set for FDK \cite{feldkamp1984practical}, Iterative \cite{Kaipio2005}, LIRE-L \cite{moriakov2022lire}, NAF \cite{zha2022naf}, and CondCBNT (ours). LIRE-L slightly outperforms CondCBNT but requires more GPU memory. Our method excels with less memory and comparable runtime. \looseness=-1}
\label{tab:data_fulltable}
    \scalebox{0.535}{
    \begin{tabular}{>{\columncolor{white}}r l c c c c c c c}
        \toprule
        &&\multicolumn{3}{c}{Noisy} & \multicolumn{3}{c}{Noise-free}&\\
        \midrule
        P. & Method & PSNR ($\uparrow$) & SSIM  ($\uparrow$) & \begin{tabular}[c]{@{}l@{}}Time \\ (s/vol) \end{tabular} & PSNR  ($\uparrow$) & SSIM ($\uparrow$) & \begin{tabular}[c]{@{}l@{}}Time \\ (s/vol) \end{tabular}& \begin{tabular}[c]{@{}l@{}}Mem. \\ (MiB) \end{tabular} \\
        \midrule
        50 & FDK  & 14.54 $\pm$ 2.90 & .20 $\pm$ .07  & 0.8 & 16.09 $\pm$ 3.22 & .43 $\pm$ .09 & 0.8 & 100 \\
        & Iterative & 26.36 $\pm$ 2.11 & .70 $\pm$ .08 & 7.7 & 27.13 $\pm$ 2.80 & .71 $\pm$ .08 & 30.8 & 300\\
        &  LIRE-L & 29.48 $\pm$ 2.07 & .83 $\pm$ .05 & 3.9 & -&-&- & 2.1k\\
        & NAF & 22.83 $\pm$ 2.24 & .58 $\pm$ .10  & 161 &24.26 $\pm$ 2.52 & .72 $\pm$ .08 & 582 & 18\\
        & \textbf{CondCBNT} & 28.31 $\pm$ 1.22 & .80 $\pm$ .05  & 124 & 30.21 $\pm$ 1.42 & .86 $\pm$ .05 & 647 & 96 \\
        \midrule
        400 & FDK & 16.43 $\pm$ 3.38 & .45 $\pm$ .12 & 7 & 16.71 $\pm$ 3.47 & .65 $\pm$ .09 & 7 & 100 \\
        & Iterative & 28.38 $\pm$ 3.27 & .78 $\pm$ .11 & 87.4 & 31.40 $\pm$ 6.22 & .91 $\pm$ .07 & 174 & 600 \\
        & LIRE-L & 30.70 $\pm$ 2.25 & .88 $\pm$ .05 & 12.8 & -&-&- & 4k\\
        & NAF & 25.93 $\pm$ 2.45 & .75 $\pm$ .08 & 275 & 25.04 $\pm$ 2.91 & .77 $\pm$ .08 & 580& 205 \\
        & \textbf{CondCBNT} & 29.89 $\pm$ 1.39 & .86 $\pm$ .05  & 763 & 30.63 $\pm$ 1.43 & .88 $\pm$ .04 & 595& 96 \\
        \bottomrule
    \end{tabular}
    }
    \end{minipage}
\end{table}
\begin{figure}
\centering \includegraphics[width=.8\linewidth]{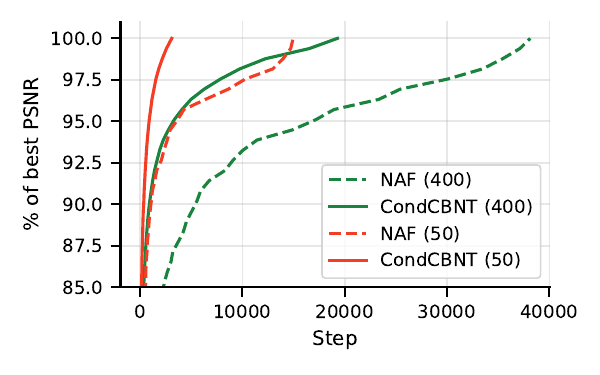}
    \captionof{figure}{Using noisy projections, the percentage of the best PSNR ($\uparrow$) that a model can reach over the number of steps required to achieve it. CondCBNT converges significantly faster.}
    \label{fig:speed_comp}
\end{figure}

\section{Experiments}

Hyperparameter search for NAF, CondCBNT, and the Iterative method was carried out on the validation set. With noisy projections, early stopping was used to avoid overfitting the noise. Instead, with noise-free projections, we decided to stop after about 10 minutes of training. Although more time would have improved performance further, it would not have provided any additional insights. It is worth noting that individual volume optimization was not conducted to reflect the constraints of a realistic scenario.

During training, we followed \citet{lin2023learning} and directly supervised the neural field with density values, as we observed this greatly improved stability. During inference on validation and test sets, we kept the shared NF fixed and only optimized the randomly initialized NMF weights for each unseen scan (see Appx. \ref{appx:experimental_details}).
We first evaluated the model on the test set using 50 and 400 noise-free projections respectively, results shown in Tab. \ref{tab:data_fulltable} right. CondCBNT greatly improves reconstruction quality both in terms of PSNR and SSIM, compared to classical methods and NAF. Next, we validated the model on 50 and 400 noisy projections, results for which are shown in Tab. \ref{tab:data_fulltable} left. Again, we see considerable improvements in our method over all baseline approaches. LIRE-L is the exception, achieving a performance slightly better than CondCBNT with significantly faster reconstruction speed at the cost of an increased memory footprint. \looseness=-1

\begin{figure}[t]
    \subfigure{
        \begin{minipage}[b]{\ratiominifigure\linewidth}
            \includegraphics[width=\linewidth]{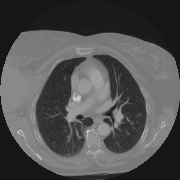}
            \includegraphics[width=\linewidth]{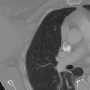}
        \end{minipage}
    }\hspace{0cm}%
    \subfigure{
        \begin{minipage}[b]{\ratiominifigure\linewidth}
            \includegraphics[width=\linewidth]{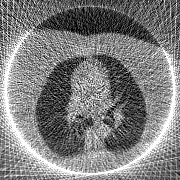}
            \includegraphics[width=\linewidth]{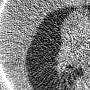}
        \end{minipage}
    }\hspace{-0.1cm}%
    \subfigure{
        \begin{minipage}[b]{\ratiominifigure\linewidth}
            \includegraphics[width=\linewidth]{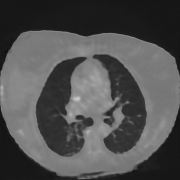}
            \includegraphics[width=\linewidth]{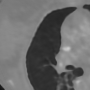}
        \end{minipage}
    }\hspace{-0.1cm}%
    \subfigure{
        \begin{minipage}[b]{\ratiominifigure\linewidth}
            \includegraphics[width=\linewidth]{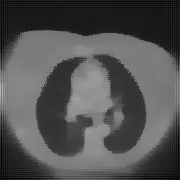}
            \includegraphics[width=\linewidth]{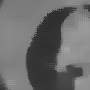}
        \end{minipage}
    }\hspace{-0.1cm}%
    \subfigure{
        \begin{minipage}[b]{\ratiominifigure\linewidth}
            \includegraphics[width=\linewidth]{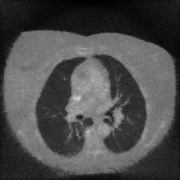}
            \includegraphics[width=\linewidth]{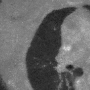}
        \end{minipage}
    }\hspace{-0.1cm}%
    \subfigure{
        \begin{minipage}[b]{\ratiominifigure\linewidth}
            \includegraphics[width=\linewidth]{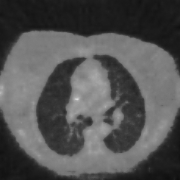}
            \includegraphics[width=\linewidth]{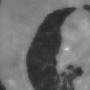}
        \end{minipage}
    }
    \vspace{0.2cm}
    \subfigure[\tiny Ground Truth]{
        \begin{minipage}[b]{\ratiominifigure\linewidth}
            \includegraphics[width=\linewidth]{figures/volumes_figures/axial/volume_242_gt.png}
            \includegraphics[width=\linewidth]{figures/volumes_figures/axial/zoom_volume_242_gt.png}
        \end{minipage}
    }\hspace{0cm}%
    \subfigure[\tiny FDK]{
        \begin{minipage}[b]{\ratiominifigure\linewidth}
            \includegraphics[width=\linewidth]{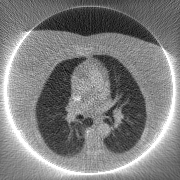}
            \includegraphics[width=\linewidth]{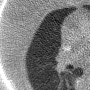}
        \end{minipage}
    }\hspace{-0.1cm}%
    \subfigure[\tiny LIRE-L]{
        \begin{minipage}[b]{\ratiominifigure\linewidth}
            \includegraphics[width=\linewidth]{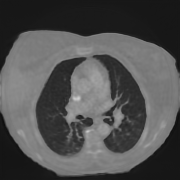}
            \includegraphics[width=\linewidth]{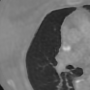}
        \end{minipage}
    }\hspace{-0.1cm}%
    \subfigure[\tiny Iterative]{
        \begin{minipage}[b]{\ratiominifigure\linewidth}
            \includegraphics[width=\linewidth]{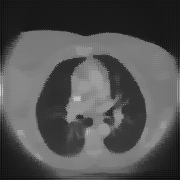}
            \includegraphics[width=\linewidth]{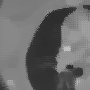}
        \end{minipage}
    }\hspace{-0.1cm}%
    \subfigure[\tiny NAF]{
        \begin{minipage}[b]{\ratiominifigure\linewidth}
            \includegraphics[width=\linewidth]{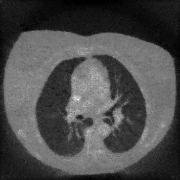}
            \includegraphics[width=\linewidth]{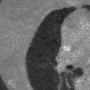}
        \end{minipage}
    }\hspace{-0.1cm}%
    \subfigure[\tiny \textbf{CondCBNT}]{
        \begin{minipage}[b]{\ratiominifigure\linewidth}
            \includegraphics[width=\linewidth]{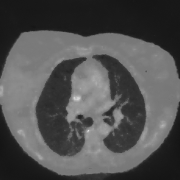}
            \includegraphics[width=\linewidth]{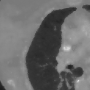}
        \end{minipage}
    }%
    \caption{Ground truth and reconstructions using all the methods applied to noisy projections. Top 50, bottom 400 projections. Grayscale with density in $[0-0.04]$. Our method does not overfit the noise and maintains tissue contrast. High-res in Appx. \ref{appx:additional_results}.}
    \label{fig:recons}
\end{figure}

Qualitative assessment in the noisy case is possible from Fig. \ref{fig:recons}, where it is evident that NAF overfits the noise. The iterative method over-smooths the reconstruction and exhibits blocky artifacts. The FDK reconstruction suffers from artifacts caused by the detector size, noise, and the low number of projections. LIRE-L and CondCBNT both reconstruct the volume with better soft-tissue contrast and without overfitting the noise.

Comparing convergence speed from Tab. \ref{tab:data_fulltable} is hard because of diverging implementation choices and final performance reached. Therefore, we normalized performance by maximum PSNR reached after optimization. Additionally, given that dataset and batch size were the same, we decided to compare using the number of iterations instead of wall-clock time (Fig \ref{fig:speed_comp}). This shows how CondCBNT quickly reaches a satisfying performance with both noisy and noise-free projections. Especially interesting is that, in the 400 projection case, CondCBNT was optimized for only half of a full epoch and still managed to outperform NAF and be within $1$ standard deviation of LIRE-L.
Since our method does not require training the whole model from scratch for a newly obtained set of projections, the model converges considerably faster.\looseness=-1

\section{Conclusion}
We improve noise resistance of neural field (NF)-based CBCT reconstruction methods by sharing a conditional NF over scans taken from different patients. We propose learning a continuous, local conditioning function expressed through a sample-specific \textit{Neural Modulation Field} which modulates activations in the conditional NF to express volume-specific details. \textit{Conditional Cone-Beam Neural Tomography} (CondCBNT) represents an efficient improvement over previous approaches, in terms of GPU memory scalability and reconstruction quality on both noise-free and noisy data and with varying numbers of available projections. \looseness=-1

\bibliography{paper_neural_tomography}
\bibliographystyle{synsml2023}

\newpage
\appendix
\onecolumn
\section{Multiresolution Hash Encoding} \label{appx:hashencoding}
\citet{muller2022instant} propose a multi-resolution hash encoding as coordinate embedding for neural fields. Here, we briefly describe the method in more detail.

Multi-resolution hash encoding is a parametric embedding, meaning the embedding function itself contains additional trainable parameters. In multi-resolution hash encoding this is done through assigning freely trainable weights to grid points from a set of multi-resolution grids defined over the input space. These parameters are then looked up and interpolated for a specific input coordinate $\mathbf{x}$. Formally, the embedding consists of a number of levels $L$, which correspond to the multiple grid resolutions, a feature dimensionality $d$ denoting the dimensionality of each trainable vector attached at a grid point, a base resolution denoting the number of grid points for the lowest resolution grid, a per-level resolution increase factor $r$ and a maximum hash-table size.

\begin{figure}
    \centering    \includegraphics[width=0.75\textwidth]{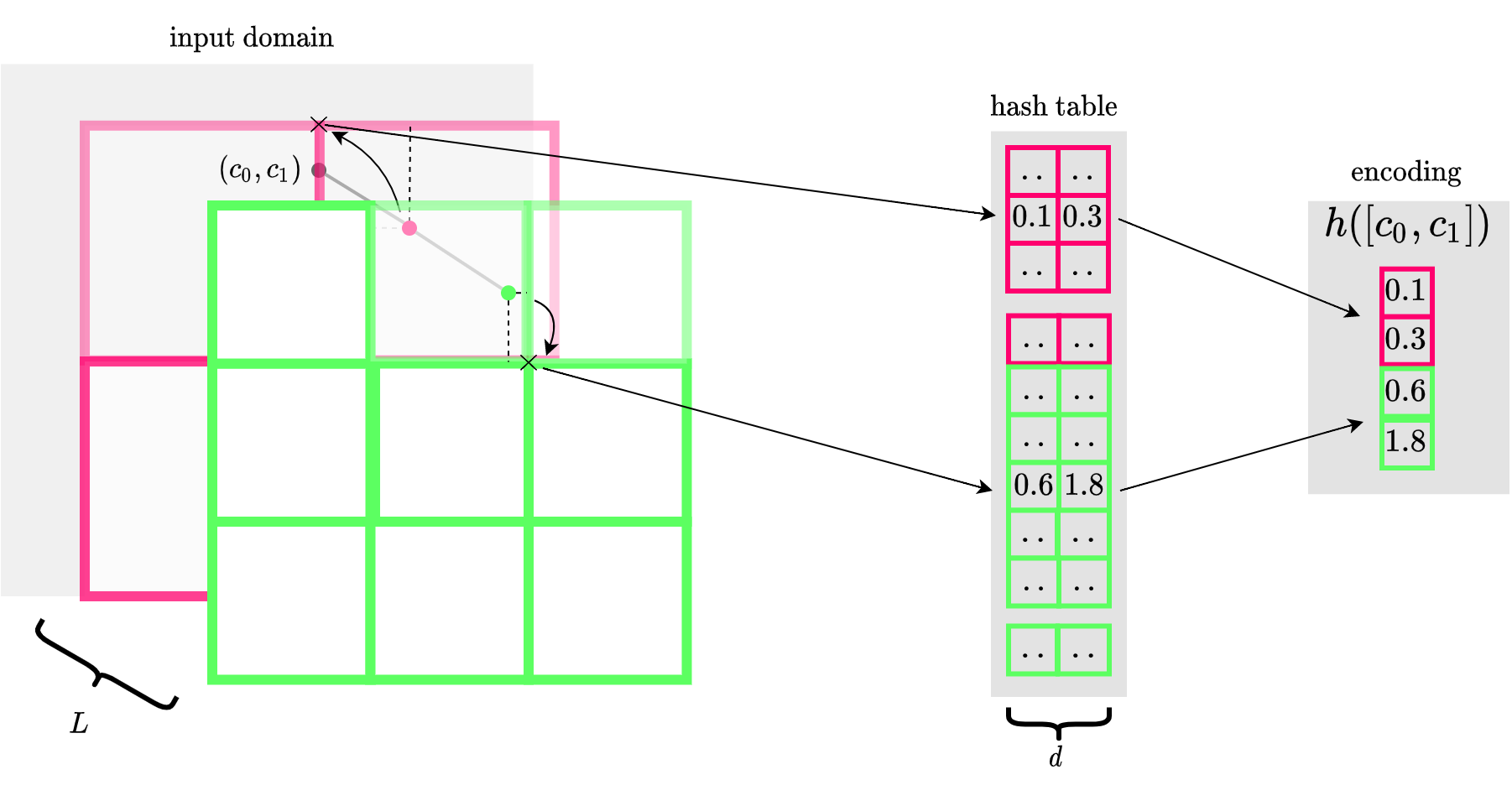}
    \caption{Multi-resolution hash encoding. $L$ grids of multiple resolutions (green, red) are defined over the input domain. Each grid-point corresponds to an entry in a hash table. Each entry in this hash table consists of a $d-$dim freely trainable weight vector. To encode a coordinate $(c_0, c_1)$, it is mapped to its closest grid-points on every grid (in practice a linear interpolation of the $n$ nearest grid points is taken), and the encoding for this coordinate is given by concatenating the grid points corresponding feature vectors, to end up with a $(L\cdot d)$- dimensional embedding.}
    \label{fig:hash_encoding}
\end{figure}

To encode a coordinate $(c_0, c_1)$, it is mapped to its closest grid-points on every grid (in practice a linear interpolation of the $n$ nearest grid points is taken), and the encoding for this coordinate is given by concatenating the grid points corresponding feature vectors, to end up with a $(L\cdot d)$- dimensional embedding. See figure \ref{fig:hash_encoding} for an illustration.

\section{Metrics}
The metrics used to quantitatively evaluate the reconstruction quality are the Peak Signal to Noise Ratio (PSNR) and the Structural Similarity Index Measure (SSIM). Both metrics can be adapted in the 3D setting in a straightforward way. Given two volumes $x, y\in \mathbb{R}^{H\times W\times D}$ where $H, W$, and $D$ are the height, width, and depth of the volume respectively, $y$ is the ground truth and $x$ is the reconstruction, the PSNR is the following
\begin{align}
\text{PSNR} (x, y) &= 10 \cdot \log_{10} \cfrac{{\left(\max {y}\right)^2}}{\text{MSE}(x, y)}\\
&= 20 \cdot \log_{10} {\max {y}} - 10 \cdot \log_{10}  \text{MSE}(x, y),
\end{align}

where the second step improves numerically stability and the MSE is the voxel-wise Mean Squared Error:
\begin{equation}
    \text{MSE}(x,y) = \cfrac{1}{N M L} \sum_{i=0}^{N-1} \sum_{j=0}^{M-1} \sum_{k=0}^{L-1} (x(i, j, k) - y(i, j, k))^2.
\end{equation}
The SSIM is computed over a small $K\times K \times K$ cube within the volume. This is repeated for all pixels, padding when necessary with zeros. Here we show the formula for the entire volume, although the original definition is for a single region:
\begin{equation*}
\hbox{SSIM}(x,y) = \frac{(2\mu_x\mu_y + c_1)(2\sigma_{xy} + c_2)}{(\mu_x^2 + \mu_y^2 + c_1)(\sigma_x^2 + \sigma_y^2 + c_2)},
\end{equation*}
where $\mu$ indicates the mean, $\sigma$ the covariance, $c_1 = (k_1L)^2$ and $c_2=(k_2L)^2$ with $k_1 = 0.01$, $k_2 = 0.03$, and $L$ the difference between the maximum value and minimum value in $y$.

\section{Experimental details} \label{appx:experimental_details}
When training NAF and CondCBNT, rays are sampled at random to form a batch. Then, a number of samples are selected along the ray to form the inputs of the model. While in NAF the batch is created using rays sampled at random from a single projection, for ConCBNT we sampled rays from any projection.

\paragraph{Projection noise} was added using the Poisson distribution, to simulate the effect of measurement noise. This is also called \textit{shot noise}, and it happens in all devices which measures the amount of photons that hit them. The probability of detecting photons can be modeled using a Poisson distribution. Intuitively, a thicker and denser substance in the path of the ray will result in a lower probability of detection and more noise in the projection. To be specific, assuming a projected value of $p$ and a fixed photon count $\pi$ (set at $5 \times 10^{5}$ in our experiments), the Poisson distribution's rate is defined as $\lambda = \pi e^{-p}$. Thus, the probability of detecting a specific number of photons, $q$, can be expressed as:

\begin{equation}
P(q;\lambda) = \frac{e^{-\lambda}\lambda^q}{q!} = \frac{(\pi e^{-p})^q e^{-\pi e^{-p}}}{q!}
\end{equation}

By sampling a value $q$ from this distribution, the resulting projected value is then calculated as:

\begin{equation}
\tilde{p} = -\log\left(\frac{q}{\pi}\right)
\end{equation}

\subsection{Architectural details}
Here, we describe the architectural specifications for the shared neural field and the patient-specific modulation neural fields. 

The shared neural field $f_{\theta}$ consists of a multi-resolution hash encoding, as described in \ref{appx:hashencoding}, with 16 levels of feature dimensionality 2, a base resolution of $16\times16\times16$, a per-level resolution increase factor of 2, and a hash-table with maximum size of $2^{19}$ parameters per level. This results in a 32-dimensional embedding, which is passed through 2 linear layers with hidden size 128, each followed by patient-specific FiLM modulation - as described in Sec. \ref{par:conditioning} - and ReLU activations. Each modulation neural field $\varphi_{\theta_i}$ also uses multi-resolution hash encoding to embed the input coordinate, followed by 2 linear layers of hidden dimensionality 128 with ReLU activations, which outputs into a $2\cdot 128$ dimensional code $\boldsymbol{z}$ split into $\boldsymbol{\gamma}, \boldsymbol{\beta}\in \mathbb{R}^{128}$.

\subsection{Hyperparameters}
In this section, we describe the hyperparameters used in the experiments. for all experiments, the code was implemented in PyTorch \cite{PyTorch}, optimized using Adam \cite{KingmaB14} with $\beta_1 = 0.9, \beta_2 = 0.999, \epsilon= 10^{-8}$.

\paragraph{CondCBNT} was trained for 15 hours on an A100 GPU using all 200 volumes from the training set. The learning rate used for the MNF was $10^{-4}$ while $10^{-3}$ for the shared NF. During training the batch size was $16,384$. During validation and testing, the MNFs are optimized individually for each patient, with a batch size of $1024$ rays and $300$ samples along the ray. We sample only points within the bounding box of the patient, defined by the original CT scan.

\paragraph{NAF} was optimized on each volume individually, with a learning rate of $5\times 10^{-4}$, optimized through hyperparameter search on the validation set. For the noise-free projection settings, the model used reflected the specifications from the original paper. The hash encoding used a base resolution of $16$, the maximum size of the hash table was $2^{21}$, the number of levels was $16$ and the size of the feature vector for each level was $2$. Instead, validation revealed that a base resolution of $8$, with $8$ levels and a hash table size of $2^{19}$ resulted in better reconstruction, as it avoided overfitting to the noise more often. For both settings, an MLP with LeakyReLU activations, 4 layers, and 32 neurons per layer was used. The batch size is also $1024$ rays, with $300$ points sampled per ray.

\section{Additional experimental results}
\label{appx:additional_results}
\subsection{Larger-scale images of reconstructed volumes}
\begin{figure*}
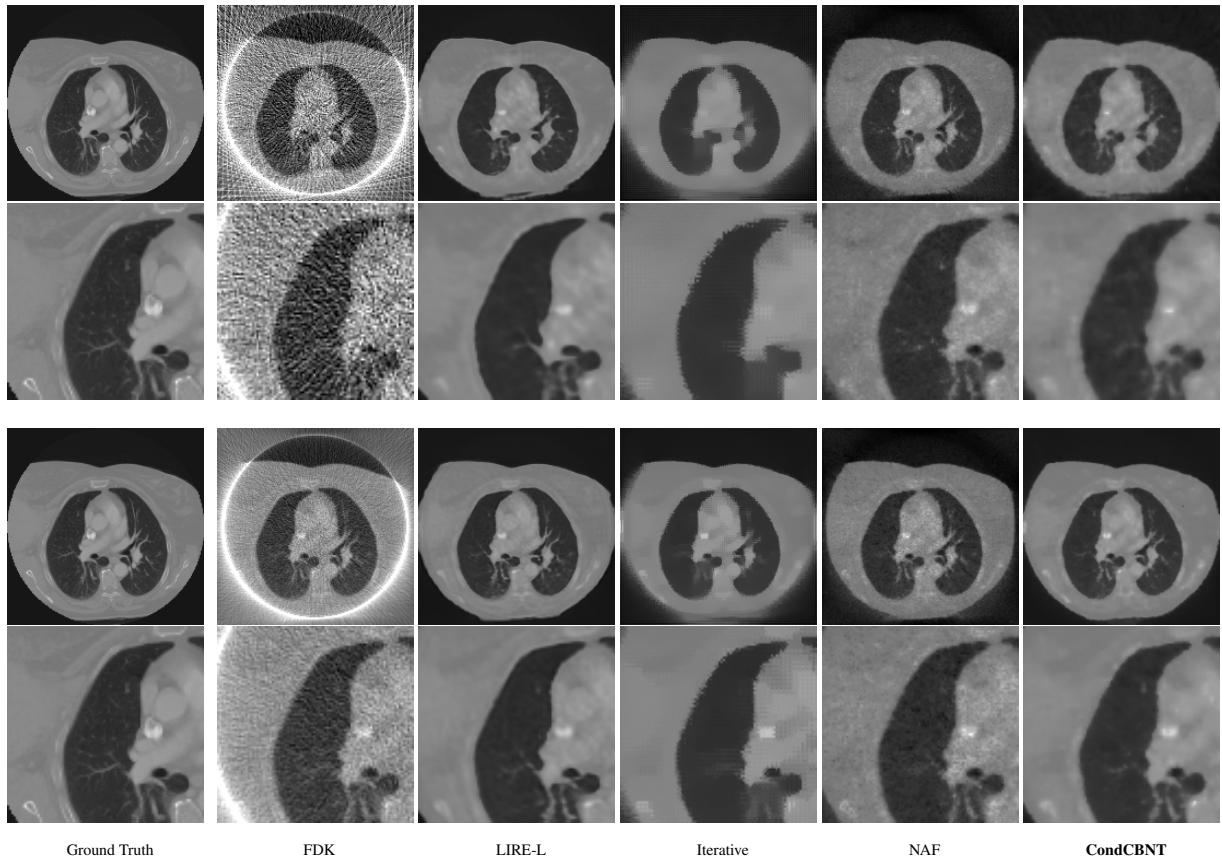

    \centering
    \subfigure{
        \begin{minipage}[b]{\ratiominifigure\linewidth}
            \includegraphics[width=\linewidth]{figures/volumes_figures/axial/volume_242_gt.png}
            \includegraphics[width=\linewidth]{figures/volumes_figures/axial/zoom_volume_242_gt.png}
        \end{minipage}
    }\hspace{0cm}%
    \subfigure{
        \begin{minipage}[b]{\ratiominifigure\linewidth}
            \includegraphics[width=\linewidth]{figures/volumes_figures/axial/volume_242_fbp_50.png}
            \includegraphics[width=\linewidth]{figures/volumes_figures/axial/zoom_volume_242_fbp_50.png}
        \end{minipage}
    }\hspace{-0.1cm}%
    \subfigure{
        \begin{minipage}[b]{\ratiominifigure\linewidth}
            \includegraphics[width=\linewidth]{figures/volumes_figures/axial/volume_242_lirel_50.png}
            \includegraphics[width=\linewidth]{figures/volumes_figures/axial/zoom_volume_242_lirel_50.png}
        \end{minipage}
    }\hspace{-0.1cm}%
    \subfigure{
        \begin{minipage}[b]{\ratiominifigure\linewidth}
            \includegraphics[width=\linewidth]{figures/volumes_figures/axial/volume_242_iter_50.png}
            \includegraphics[width=\linewidth]{figures/volumes_figures/axial/zoom_volume_242_iter_50.png}
        \end{minipage}
    }\hspace{-0.1cm}%
    \subfigure{
        \begin{minipage}[b]{\ratiominifigure\linewidth}
            \includegraphics[width=\linewidth]{figures/volumes_figures/axial/naf_400_volume_242.png}
            \includegraphics[width=\linewidth]{figures/volumes_figures/axial/zoom_naf_400_volume_242.png}
        \end{minipage}
    }\hspace{-0.1cm}%
    \subfigure{
        \begin{minipage}[b]{\ratiominifigure\linewidth}
            \includegraphics[width=\linewidth]{figures/volumes_figures/axial/nefcbct_50_volume_242.png}
            \includegraphics[width=\linewidth]{figures/volumes_figures/axial/zoom_nefcbct_50_volume_242.png}
        \end{minipage}
    }
    \vspace{0.2cm}
    \subfigure[\tiny Ground Truth]{
        \begin{minipage}[b]{\ratiominifigure\linewidth}
            \includegraphics[width=\linewidth]{figures/volumes_figures/axial/volume_242_gt.png}
            \includegraphics[width=\linewidth]{figures/volumes_figures/axial/zoom_volume_242_gt.png}
        \end{minipage}
    }\hspace{0cm}%
    \subfigure[\tiny FDK]{
        \begin{minipage}[b]{\ratiominifigure\linewidth}
            \includegraphics[width=\linewidth]{figures/volumes_figures/axial/volume_242_fbp_400.png}
            \includegraphics[width=\linewidth]{figures/volumes_figures/axial/zoom_volume_242_fbp_400.png}
        \end{minipage}
    }\hspace{-0.1cm}%
    \subfigure[\tiny LIRE-L]{
        \begin{minipage}[b]{\ratiominifigure\linewidth}
            \includegraphics[width=\linewidth]{figures/volumes_figures/axial/volume_242_lirel_400.png}
            \includegraphics[width=\linewidth]{figures/volumes_figures/axial/zoom_volume_242_lirel_400.png}
        \end{minipage}
    }\hspace{-0.1cm}%
    \subfigure[\tiny Iterative]{
        \begin{minipage}[b]{\ratiominifigure\linewidth}
            \includegraphics[width=\linewidth]{figures/volumes_figures/axial/volume_242_iter_400.png}
            \includegraphics[width=\linewidth]{figures/volumes_figures/axial/zoom_volume_242_iter_400.png}
        \end{minipage}
    }\hspace{-0.1cm}%
    \subfigure[\tiny NAF]{
        \begin{minipage}[b]{\ratiominifigure\linewidth}
            \includegraphics[width=\linewidth]{figures/volumes_figures/axial/naf_50_volume_242.png}
            \includegraphics[width=\linewidth]{figures/volumes_figures/axial/zoom_naf_50_volume_242.png}
        \end{minipage}
    }\hspace{-0.1cm}%
    \subfigure[\tiny \textbf{CondCBNT}]{
        \begin{minipage}[b]{\ratiominifigure\linewidth}
            \includegraphics[width=\linewidth]{figures/volumes_figures/axial/nefcbct_400_volume_242.png}
            \includegraphics[width=\linewidth]{figures/volumes_figures/axial/zoom_nefcbct_400_volume_242.png}
        \end{minipage}
    }%
    
    \caption{Ground truth and reconstructions using all the methods applied to noisy projections. Top 50 projections, bottom 400 projections. Grayscale colormap with density in $[0-0.04]$. The detector size causes a dense ring to appear in the FDK reconstruction. NAF overfits the noise with both 50 and 400 projections. Iterative over-smooths the soft tissues and removes bones. LIRE-L succeeds in keeping soft-tissue contrast and reconstructing bones. Our method succeeds in not overfitting the noise and maintaining higher tissue constrast.}
    \label{fig:recons-but-bigger}
\end{figure*}

\begin{figure*}
    \centering
    \subfigure{
        \begin{minipage}[b]{\ratiominifigure\linewidth}
            \includegraphics[width=\linewidth]{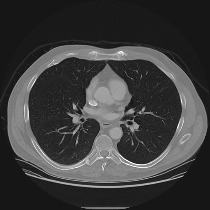}
            \includegraphics[width=\linewidth]{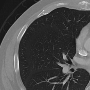}
        \end{minipage}
    }\hspace{0cm}%
    \subfigure{
        \begin{minipage}[b]{\ratiominifigure\linewidth}
            \includegraphics[width=\linewidth]{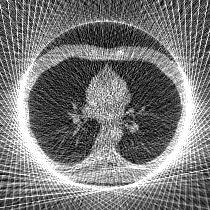}
            \includegraphics[width=\linewidth]{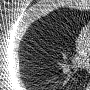}
        \end{minipage}
    }\hspace{-0.1cm}%
    \subfigure{
        \begin{minipage}[b]{\ratiominifigure\linewidth}
            \includegraphics[width=\linewidth]{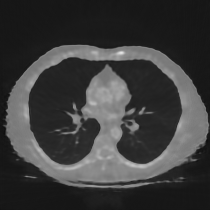}
            \includegraphics[width=\linewidth]{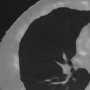}
        \end{minipage}
    }\hspace{-0.1cm}%
    \subfigure{
        \begin{minipage}[b]{\ratiominifigure\linewidth}
            \includegraphics[width=\linewidth]{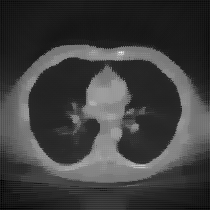}
            \includegraphics[width=\linewidth]{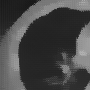}
        \end{minipage}
    }\hspace{-0.1cm}%
    \subfigure{
        \begin{minipage}[b]{\ratiominifigure\linewidth}
            \includegraphics[width=\linewidth]{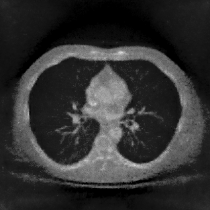}
            \includegraphics[width=\linewidth]{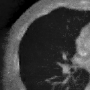}
        \end{minipage}
    }\hspace{-0.1cm}%
    \subfigure{
        \begin{minipage}[b]{\ratiominifigure\linewidth}
            \includegraphics[width=\linewidth]{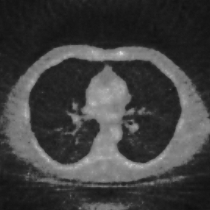}
            \includegraphics[width=\linewidth]{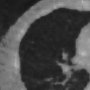}
        \end{minipage}
    }
    \vspace{0.2cm}
    \subfigure[\tiny Ground Truth]{
        \begin{minipage}[b]{\ratiominifigure\linewidth}
            \includegraphics[width=\linewidth]{figures/volumes_figures/axial/volume_236_gt.png}
            \includegraphics[width=\linewidth]{figures/volumes_figures/axial/zoom_volume_236_gt.png}
        \end{minipage}
    }\hspace{0cm}%
    \subfigure[\tiny FDK]{
        \begin{minipage}[b]{\ratiominifigure\linewidth}
            \includegraphics[width=\linewidth]{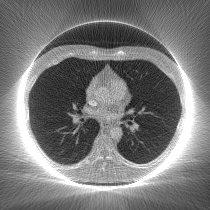}
            \includegraphics[width=\linewidth]{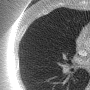}
        \end{minipage}
    }\hspace{-0.1cm}%
    \subfigure[\tiny LIRE-L]{
        \begin{minipage}[b]{\ratiominifigure\linewidth}
            \includegraphics[width=\linewidth]{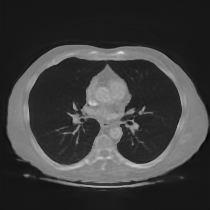}
            \includegraphics[width=\linewidth]{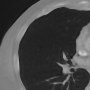}
        \end{minipage}
    }\hspace{-0.1cm}%
    \subfigure[\tiny Iterative]{
        \begin{minipage}[b]{\ratiominifigure\linewidth}
            \includegraphics[width=\linewidth]{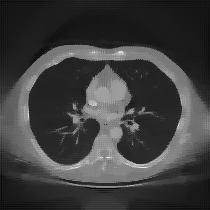}
            \includegraphics[width=\linewidth]{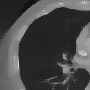}
        \end{minipage}
    }\hspace{-0.1cm}%
    \subfigure[\tiny NAF]{
        \begin{minipage}[b]{\ratiominifigure\linewidth}
            \includegraphics[width=\linewidth]{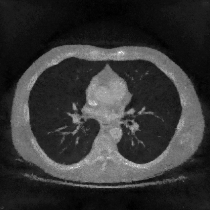}
            \includegraphics[width=\linewidth]{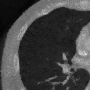}
        \end{minipage}
    }\hspace{-0.1cm}%
    \subfigure[\tiny \textbf{CondCBNT}]{
        \begin{minipage}[b]{\ratiominifigure\linewidth}
            \includegraphics[width=\linewidth]{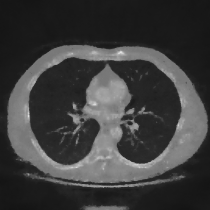}
            \includegraphics[width=\linewidth]{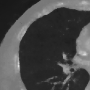}
        \end{minipage}
    }%
    
    \caption{Ground truth and reconstructions using all the methods applied to noisy projections. Top 50 projections, bottom 400 projections. Grayscale colormap with density in $[0-0.04]$. Similar behavior to the one shown in Fig. \ref{fig:recons-but-bigger}. Soft-tissue contrast resolution very clear for CondCBNT and LIRE-L, thanks to less noise in the projections. NAF still overfits the noise. Less over-smoothing by the iterative method.}
    \label{fig:recons-but-bigger-236}
\end{figure*}

For an improved viewing experience, we include larger-scale versions of our experimental results in Fig. \ref{fig:recons-but-bigger} and Fig. \ref{fig:recons-but-bigger-236}. The latter shows a volume with less noise in the projections.

\section{Acknowledgments}
The authors acknowledge the National Cancer Institute and the Foundation for the National Institutes of Health, and their critical role in the creation of the free publicly available LIDC/IDRI Database used in this study.

\end{document}